\documentclass[epj]{svjour}

\usepackage{amsmath} 
\usepackage{amssymb} 
\usepackage{graphicx} 

\newcommand{\bsigma}{\mbox{\boldmath $\sigma$}} 

\newcommand{\btau}{\mbox{\boldmath $\tau$}}

\newcommand{\hach}{h_{i}(\bsigma)}

\newcommand{\hachtt}{h_{i}(\bsigma(t+1))}

\newcommand{\betinf}{\beta\rightarrow\infty}

\begin{document}
\title{Two-cycles in spin-systems:  multi-state Ising-type ferromagnets}
\author{D. Boll\'e \and J. Busquets Blanco 
}                     
%
%
\institute{Instituut voor Theoretische Fysica, 
                 Katholieke Universiteit Leuven, 
                 Celestijnenlaan 200 D, 
                 B-3001 Leuven, Belgium }
\date{Received: date / Revised version: date}
%
\abstract{Hamiltonians for general multi-state spin-glass systems with       
Ising symmetry are derived for both sequential and synchronous updating of the
spins. The possibly different behaviour caused by the way of updating is studied
in detail for the (anti)-ferromagnetic version of the models, which can be 
solved analytically without any approximation, both thermodynamically via a 
free-energy calculation and dynamically using the generating functional 
approach. Phase diagrams are discussed and the appearance of two-cycles in the
case of synchronous updating is examined. A comparative study is made for the
$Q$-Ising and the 
Blume-Emery-Griffiths ferromagnets and some interesting physical differences 
are found. Numerical simulations confirm the results obtained.
} 
\maketitle
\section{Introduction}
\label{intro}
Recently, there has been renewed interest in the possibly different physics 
arising from the sequential or synchronous execution of the microscopic update  
rule of the spins in disordered systems. For example, the Little- Hopfield model
on a random graph \cite{PS03} and random field Ising chains \cite{SC00}, both 
with synchronous updating have been studied. For both systems governed by a
pseudo- Hamiltonian of binary Ising spins (i.e., a Hamiltonian dependent on the
inverse temperature) first derived by Peretto \cite{P84}, it has been shown
that the physics is asymptotically identical to that of the sequential version
of the models. Furthermore, for a class of attractor neural networks with 
spatial structure (one dimensional nearest-neighbour interactions and
infinite-range interactions) governed again by Peretto's pseudo-Hamiltonian 
it has been found \cite{SC200} that dynamical transition lines for synchronous
updating in parameter space are exact reflections in the origin of those in
sequential updating and that the relevant macroscopic observables can be
obtained from those of sequential updating via simple transformations. 

It is yet unclear to what extent the two types of spin updating
lead to such common equilibrium features. For example, it is
known that the phase diagram of the sequential and synchronous Hopfield 
neural network model in the replica-symmetric approximation are different (e.g.
the retrieval region is slightly larger in the synchronous case) \cite{FK88} 
whereas the phase diagram
of the Sherrington-Kirkpatrick (SK) spin-glass model \cite{SK72} remains 
unaffected by this difference in updating \cite{Nreport}.

The aim of this work is to get more insight in the possible differences between
sequential and synchronous updating by studying more complicated models
containing multi-state spins. In particular, we look at the $Q$-Ising and
Blume-Emery-Griffiths (BEG) (anti)-ferromagnetic models which can also be
related to a neural network model storing 
only one pattern. We discuss the relevant phase 
diagrams  as well as the dynamics using the generating functional approach. We
are not aware of any previous studies comparing these two types of updating 
for these models. It turns out that for the $Q$-Ising model again the  
transition lines for synchronous updating in parameter space are exact 
reflections of those in sequential updating. For the BEG model, however, the
two forms of updating lead to different physics in part of the parameter space.

The paper is organized as follows. In section \ref{sec2}, we generalize the
results of Peretto  by writing down (pseudo) Hamiltonians based on the detailed 
balance property for multi-state spin models with random interactions for both 
sequential and synchronous updating. In the zero- temperature limit  
the corresponding Lyapunov functions are obtained. In section \ref{sec3}, the
phase diagrams for the 
$Q$-Ising  (anti)-ferromagnets are studied in detail, emphasizing the 
differences between both forms of spin updating. Section~\ref{sec4} discusses
the statics of the BEG (anti)-ferromagnet and section~\ref{sec5} the dynamics 
using the generating functional 
approach. Numerical simulations confirm the results obtained.
Finally, in section \ref{sec6} some concluding remarks are presented.

\section{Models and Hamiltonians}
\label{sec2}
\subsection{The $Q$-Ising model}

Consider a model of $N$ spins which can take values 
$\sigma_i, i=1, \ldots ,N$ from a discrete set
        $ {\cal S} = \lbrace -1 = s_1 < s_2 < \ldots < s_Q
                = +1 \rbrace $.
Given the configuration
        ${\bsigma}(t)\equiv\{\sigma_i(t)\},
        i=1,\ldots,N$,
the local field in spin $i$ equals
\begin{equation}
        \label{eq:h}
        h_i({\bsigma}(t))=
                \sum_{j=1}^N J_{ij}\sigma_j(t)
\end{equation}
with $J_{ij}$ the interaction between spin $i$ and spin $j$.
In general, the $J_{ij}$ are quenched random variables chosen according to a 
certain distribution, e.g. a Gaussian in the case of the SK model or a Hebbian
learning rule in the case of neural networks.

A spin is updated through the spin-flip dynamics
defined by the transition probability
\begin{equation}
      \Pr \{\sigma_i(t+1) = s_k \in {\cal S} | \bsigma(t) \}
        =
        \frac
        {\exp [- \beta \epsilon_i(s_k|\bsigma(t))]}
        {\sum_{s \in {\cal S}} \exp [- \beta \epsilon_i
                                   (s|\bsigma(t))]}\,.
\label{eq:trans}
\end{equation}
Here the energy potential $\epsilon_i[s|{\bsigma}]$ is defined 
by \cite{R90}
\begin{equation}
        \epsilon_i[s|{\bsigma}]=
                -\left[\frac{1}{2}h_i({\bsigma})s-bs^2\right]
         \,, \label{eq:energy}
\end{equation}
where $b$ is the gain parameter of the system.
The zero temperature limit $T=\beta^{-1} \rightarrow 0$ of this dynamics
is given by the updating rule
\begin{equation}
        \label{eq:enpot}
        \sigma_i(t)\rightarrow\sigma_i(t+1)=s_k:
                \min_{s\in{\cal S}} \epsilon_i[s|{\bsigma}(t)]
            =\epsilon_i[s_k|{\bsigma}(t)]
\,.
\end{equation}
This updating rule (\ref{eq:enpot}) is equivalent to using a gain 
function $\mbox{g}_b(\cdot)$,
\begin{eqnarray}
        \label{eq:gain}
        &&\sigma_i(t+1)   =   
               \mbox{g}_b(h_{i}(\bsigma(t))
                  \nonumber      \\
               &&\mbox{g}_b(x) \equiv \sum_{k=1}^Qs_k
                        \left[\theta\left[b(s_{k+1}+s_k)-x\right]-
                              \theta\left[b(s_k+s_{k-1})-x\right]
                        \right] \nonumber \\
\end{eqnarray}
with $s_0\equiv -\infty$ and $s_{Q+1}\equiv +\infty$. The parameter $b$ 
suppresses or enhances the states of the spins that lie around the zero state.

In the case of sequential updating it is well-known
from detailed balance arguments  that for symmetric couplings,
i.e., $J_{ij}= J_{ji}$, and in the absence of self-interactions, i.e., 
$J_{ii}=0$, the equilibrium distribution for the $Q$-Ising system has the 
Boltzmann form with  Hamiltonian (see, e.g., \cite{GS77})
\begin{equation}
H_S(\bsigma)=-\frac{1}{2}
  \sum_{i,j\neq i}^NJ_{ij}\sigma_i\sigma_j+b\sum_{i=1}^N\sigma_i^2\, ,
\label{QIseq}
\end{equation}
valid for any temperature and with Lyapunov behaviour for $T=0$. 
We remark that the second term is not a self coupling term. As in any
spin model with sequential updating, the
stationary solutions can only be fixed points.

For synchronous updating a discussion does not seem to have appeared in the
literature. In that case, the arguments of Peretto \cite{P84} can be
generalized rather straightforwardly to obtain that again the equilibrium 
probability distribution can be written in the Boltzmann form with a 
Hamiltonian dependent on the inverse temperature
\begin{equation}
H_P(\bsigma)=-\frac{1}{\beta}\sum_{i=1}^N\ln{\left[\sum_{s\in\mathcal{S}}
                \exp{(\beta[h_i(\bsigma)s-bs^2])}\right]}
                 +b\sum_{i=1}^N\sigma_i^2
\label{parQham}
\end{equation}
We remark that  self-couplings $J_{ii}$ are allowed to be present.
This pseudo-Hamiltonian can be written in a two-spin representation 
\begin{eqnarray}
H_P(\bsigma,\btau)
&=&-\frac{1}{\beta}\sum_{i,j\neq
   i}J_{ij}\sigma_i\tau_j+b\sum_i(\sigma_i^2+\tau_i^2) 
   \label{QIsyn}\\
&=&-\frac{1}{\beta}\sum_{i,j\neq i}J_{ij}\sigma_i(t)\sigma_j(t+1) 
      \nonumber \\
 &+& b \sum_i[\sigma_i^2(t)+\sigma_i^2(t+1)]
\end{eqnarray}
In the limit $\betinf$ we find after some algebra starting from (\ref{parQham})
\begin{eqnarray}
\hspace*{-1cm} H_{P}(\bsigma;T=0) \nonumber \\
 && \hspace*{-1.5cm}
  =-\sum_{i=1}^{N}\sum^{Q}_{j=[\frac{Q+3}{2}]}\left(|\hach|s_j-bs_j^2\right) 
     \nonumber\\
  &&  \times  \theta\left(|\hach|-b(s_{j-1}+s_j)\right) \nonumber \\
   &&  \times      \theta\left(b(s_{j}+s_{j+1})-|\hach|\right)
                    +b\sum_{i=1}^{N}\sigma_i^2
\end{eqnarray}
with the standard notation $[\cdot]$ indicating the largest integer. 
For $Q=2$, we find back the Hopfield Hamiltonian with an irrelevant additive 
constant. For $Q=3$ we have 
\begin{equation}
H_{P}(\bsigma;T=0)=
     -\sum_{i=1}^N\left(|\hach|-b\right)\theta\left(|\hach|-b\right)
                         +b\sum_{i=1}^N\sigma_i^2
\end{equation}

The Hamiltonian for general $Q$ is  bounded from below by
$H_{P}(\bsigma;T=0)\geq-\sum_{i,j\neq i}|J_{ij}|-N|b|$ and, furthermore
\begin{eqnarray}
\Delta H_{P}(\bsigma;T=0) 
&&\equiv H_{P}(\bsigma(t+1);T=0)-H_{P}(\bsigma(t);T=0) \nonumber \\
  &&= -\sum_{i=1}^N\left(\sigma_i(t+2)-\sigma_i(t)\right) \nonumber \\
  &&   \times   \left(\hachtt -b(\sigma_i(t+2)+\sigma_i(t))\right)\leq 0
              \nonumber \\
\end{eqnarray}  
indicating that the equilibrium behaviour can be fixed-points and/or cycles of 
period $2$, i.e., $\sigma_i(t)=\sigma_i(t+2)$,  $\forall i$.

\subsection{The BEG model}
The second model we consider is the BEG model introduced in \cite{BEG71}
in the context of the $\lambda$-transition and phase separation in the mixtures
of $He^3-He^4$ in a crystal field, and recently discussed as a spin-glass 
(see \cite{ACN00,CL04} and references therein) and as a neural network model
maximising the mutual information content for ternary neurons 
\cite{DK00,toni,B04}.
This model can be described as follows.

 Consider $N$ spins which can take values 
$\sigma_i, i=1, \ldots ,N$ from a discrete set
        $ {\cal S} = \lbrace -1,0,+1\rbrace $.
Given the configuration ${\bsigma}(t)\equiv\{\sigma_i(t)\},
i=1,\ldots,N$ at time $t$, the spins are updated according to the spin-flip 
dynamics defined by the transition probability (\ref{eq:trans})
where the energy potential $\epsilon_i[s|{\bsigma}(t)]$ is now
defined by 
\begin{equation} 
          \epsilon_i[s|{\bsigma}(t)] =
          -(h_{1,i}(\bsigma(t))s+h_{2,i}(\bsigma(t))s^2) \, .
\label{potbeg} 
\end{equation} 
The local field $h_{1,i}(\bsigma)$ is the usual one as appearing, e.g.,  in 
(\ref{eq:energy}), while the biquadratic $h_{2,i}(\bsigma)$ local field is 
given by
\begin{equation} 
      h_{2,i}(\bsigma(t))=\sum_{j=1}^NK_{ij}\sigma_{j}^{2}(t) \, .
\end{equation} 
In the limit $\betinf$ this dynamics is given by the updating rule
\begin{equation}
\sigma_i(t+1)=\mbox{sign}(h_{1,i}(\bsigma(t)))
\theta\left(|h_{1,i}(\bsigma(t))|+h_{2,i}(\bsigma(t))\right)
\label{transBEG}
\end{equation}

In the case of sequential updating of the spins the Hamiltonian is known in 
the literature  mentioned above and given by 
\begin{equation}
H^{BEG}_S(\bsigma)=-\frac{1}{2}\sum_{i,j\neq
  i}^N\left(J_{ij}\sigma_i\sigma_j
           +K_{ij}\sigma_i^2\sigma_j^2\right)
\label{BEGhseq}
\end{equation}
with $J_{ii}=0$ and $J_{ij}=J_{ji}$, for $\forall i,j$.

For synchronous updating detailed balance and symmetry in the couplings lead 
to the pseudo-Hamiltonian
\begin{equation}
H_P^{BEG}(\bsigma)=
    -\frac{1}{\beta}\sum_{i=1}^{N}\ln{\left\{2e^{\beta h_{2,i}(\bsigma)}
                    \cosh{(\beta h_{1,i}(\bsigma))}+1\right\}}
\label{BEGhsyn}
\end{equation}
with as two-spin representation
\begin{eqnarray}
&& \hspace*{-0.5cm} H_P^{BEG}(\bsigma,\btau) \nonumber \\
&&\hspace*{-0.5cm} 
     =-\sum_{i,j\neq  i}\left(J_{ij}\sigma_i\tau_j
           +K_{ij}\sigma_i^2\tau_j^2\right) \\
&& \hspace*{-0.5cm}
     \equiv -\sum_{i,j\neq i}\left(J_{ij}\sigma_i(t)\sigma_j(t+1)
           +K_{ij}\sigma_i^2(t)\sigma_j^2(t+1)\right) 
\end{eqnarray}
Determining the dominant contributions in the limit $\betinf$
we find 
\begin{multline}
 H^{BEG}_{P}(\bsigma,\btau;T=0)
=-\sum_{i=1}^{N}\left(|h_{1,i}(\bsigma)|
+h_{2,i}(\bsigma)\right)\\
\times\theta\left(|h_{1,i}(\bsigma)|+h_{2,i}(\bsigma)\right)\, .
\end{multline}
This form is clearly bounded from below and it can also be shown that the
equilibrium behaviour is given by fixed-point attractors and/or cycles of period
$2$.

Both the $Q$-Ising and BEG spin-glass models and neural networks have been 
discussed in the literature starting from the Hamiltonian appropriate for 
sequential updating of the spins, as discussed in the introduction. Concerning
synchronous updating, especially concerning the appearance and properties of 
two-cycles, very little seems to be written down, even for the ferromagnetic
versions of these models. Since interesting different physics is involved 
 we want to fill this gap in the following sections.

\section{$Q$-Ising ferromagnet: sequential versus synchronous updating}
\label{sec3}

\subsection{Stationary behaviour}

We consider the $Q$-Ising (pseudo)-Hamiltonians for sequential and synchronous 
updating derived before for simplified interactions of the form 
\begin{equation}
J_{ij}=\frac{J}{N}
\end{equation}
where $J$ can be positive or negative. The parameters describing the properties
of this system are the magnetization $m$ and the spin activity $a$ given by 
\begin{equation}
   m(\bsigma)=\frac{1}{N}\sum_i\sigma_i\,, \quad 
   a(\bsigma)=\frac{1}{N}\sum_i\sigma_i^2  
\label{maqising}
\end{equation}
and in both cases the equilibrium behaviour can be studied by looking at 
the free energy per site, 
\begin{equation}
  f=\frac{-1}{\beta N}\ln{Z}\, , \quad 
  Z=\sum_{\sigma}\exp{(-\beta H(\bsigma))}. 
\end{equation}

For sequential updating starting from the Hamiltonian (\ref{QIseq}) a standard
calculation leads to the following free energy 
\begin{equation}
\beta f_{S} = \underset{m}{\mbox{extr}} \left[ \frac{\beta J}{2} m^2 
                        - \ln \sum_{\sigma} \exp (
                        {- \beta \tilde H(\sigma))}\right]
\end{equation}
with the effective Hamiltonian
\begin{equation}
\tilde H_{S}(\sigma) =  - J m \sigma  + b \sigma^2 \ .
\end{equation}
The saddle-point equation for $m$ in this notation reads
\begin{equation}
m = \frac{\sum_{\sigma}\sigma \exp({-\beta \tilde H(\sigma)})}
         {\sum_{\sigma} \exp({-\beta \tilde H(\sigma)})}
  \equiv \left\langle \sigma \right \rangle
  \label{saddlem}        
\end{equation}
which is an effective thermal average, denoted by 
 $\left\langle \cdot \right \rangle$.

In the case of synchronous updating we start from the pseudo-Hamiltonian 
written in the two-spin representation (\ref{QIsyn}).
This Hamiltonian is symmetric with respect to the transformation 
$\sigma \leftrightarrow \tau$. 
The following result for the free energy is obtained 
\begin{equation}
\beta f_{P}
  = \underset{m_{\sigma},m_{\tau}}{\mbox{extr}} 
   \left[ \beta J m_{\sigma} m_{\tau} - 
     \ln \sum_{\sigma,\tau} \exp ({-\beta \tilde
                       H(\sigma,\tau)})\right]
\end{equation}
with the effective Hamiltonian
\begin{equation}
\tilde H(\sigma,\tau) = -J m_{\tau} \sigma - J m_{\sigma} \tau 
      + b \sigma^2 + b \tau^2 \ .
\end{equation}
The saddle-point equations for $m_{\sigma}$ and $m_{\tau}$ read
\begin{eqnarray}
&&m_{\sigma} = \frac{\sum_{\sigma,\tau}\sigma 
      \exp({-\beta \tilde H(\sigma,\tau)})}
         {\sum_{\sigma,\tau} \exp({-\beta \tilde H(\sigma,\tau)})}
  = \left\langle\sigma\right\rangle     
\label{eq:intro:magn-cwQp1} \\
&&m_{\tau} = \frac{\sum_{\sigma,\tau}\tau 
      \exp({-\beta \tilde H(\sigma,\tau)})}
         {\sum_{\sigma,\tau} \exp({-\beta \tilde H(\sigma,\tau)})}
  = \left\langle\tau\right\rangle
        \ ,
\label{eq:intro:magn-cwQp}
\end{eqnarray}
where this again defines the average $\left\langle\cdot\right\rangle$.

The effective Hamiltonian factorises over the two effective spins, and so does
the partition function.  The saddle-point equations for $m_{\sigma}$ and 
$m_{\tau}$ (\ref{eq:intro:magn-cwQp1})-(\ref{eq:intro:magn-cwQp}) can be 
written as
\begin{equation}
m_{\sigma} = F_Q(m_{\tau})
\ ,
  \qquad
m_{\tau} = F_Q(m_{\sigma})
\label{saddlemFm}
\end{equation}
with the function $F_Q$ given by
\begin{equation}
    F_Q(x) = \frac{\sum_{\sigma} \sigma \exp{\beta (J x \sigma -  b \sigma^2)}}
                {\sum_{\sigma} \exp{\beta (J x \sigma -  b \sigma^2 )}}
\ .
\end{equation}
The equations (\ref{saddlemFm}) can be written as 
\begin{equation}
m_{\sigma}= F_Q(F_Q(m_{\sigma}))\, , \quad m_{\tau}= F_Q(F_Q(m_{\tau}))
   \label{parval1}  
\end{equation}
and similar equations can be written down for the activity, for instance   
\begin{align}
a_{\sigma}&= G_Q(F_Q(m_{\sigma})),\nonumber\\ 
G_Q(x)& = \frac{\sum_{\sigma} \sigma^2 
                \exp{\beta (J x \sigma -  b \sigma^2)}}
                {\sum_{\sigma} \exp{\beta (J x \sigma -  b \sigma^2 )}}\, .
   \label{parval2}
\end{align}
At this point we remark that the saddle-point equation for the sequential 
$Q$-Ising model (\ref{saddlem}) is equivalent to $m = F_Q(m)$ and the activity 
satisfies $a = G_Q(m)$.

For $J>0$ the function $F_Q(x)$ is monotonically increasing since
\begin{eqnarray}
  \frac{\partial F_Q(x)}{\partial x}
    &=& \beta J \left[ \rule{0cm}{0.6cm}
      \frac{\sum_{\sigma} \sigma^2 \, \exp{\beta (J x \sigma - b \sigma^2 )}}
      {\sum_{\sigma} \exp{\beta (J x \sigma -  b \sigma^2 )}}
      \right.  \nonumber\\
&&     -\left. \left(\frac{\sum_{\sigma} 
                 \sigma \, \exp{\beta (J x \sigma -  b \sigma^2 )}}
      {\sum_{\sigma} \exp{\beta (J x \sigma -  b \sigma^2 )}}\right)^2
      \right]
      \ge 0 \ .
\end{eqnarray}
Consequently the right-hand side of
\begin{equation}
(m_{\sigma}-m_{\tau})^2 = (m_{\sigma}-m_{\tau})(F_Q(m_{\tau}) - F_Q(m_{\sigma})) 
\end{equation}
is always negative implying that $m_{\sigma}=m_{\tau}$ and
\begin{equation}
f_{P} = 2 f_{S} \ .
\nonumber
\end{equation}
In other words, the equilibrium states for both types of updating in the
ferromagnetic $Q$-Ising model are the same. For $J<0$, and also for the 
BEG-model, this is not valid as we will see in the following sections.

\subsection{An illustrative example: $Q=3$}

\begin{figure}[h]
\resizebox{0.95\columnwidth}{!}{
  \includegraphics*[angle=270,scale=0.4]{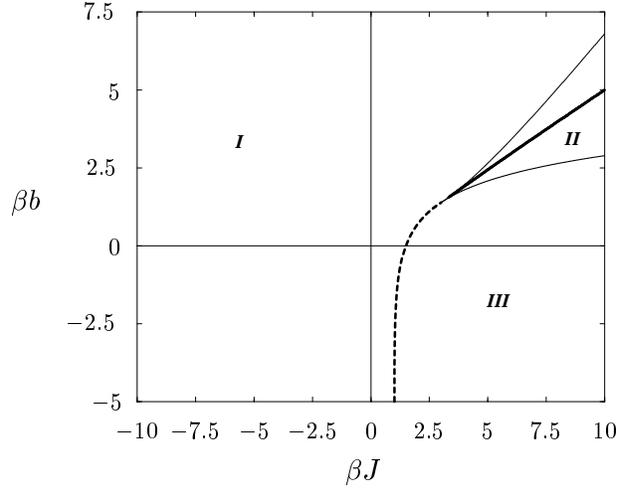}}
\caption{Phase diagram for the sequential $Q=3$ Ising ferromagnet. 
  The thick dashed (solid) line indicates the thermodynamic second (first) 
  order transition between the ferromagnet and paramagnetic phase. The 
  thin lines border the coexistence region. }
\label{Q=3_seq_pd}       
\end{figure}

The results for $Q=2$ are standard textbook knowledge (see, e.g, \cite{C01s}).
For $Q=3$, the equations for $F_Q$ and $G_Q$ can be worked out explicitly
\begin{eqnarray}
&& F_{Q=3}(x)=\frac{2\sinh{(\beta Jx)}}{\exp{(\beta b)}+
  2\cosh{(\beta Jx)}}\\
&& G_{Q=3}(x)=F_{Q=3}(x)\coth{(\beta Jx)} \, . 
\end{eqnarray}
The phase diagram for sequential updating is shown in fig.~\ref{Q=3_seq_pd}. 
It is trivial to check that a negative $J$  implies that $m=F_{Q=3}(m)$
only leads to a stable paramagnetic solution $m=0$. For positive $J$ a 
transition occurs between the paramagnetic and the ferromagnetic phase. It is 
second order and given by the dashed line
\begin{equation}
\label{cont}
a=\frac{1}{\beta J},\qquad \beta b=\ln{(2(\beta J-1))}
\end{equation}
for $\beta J < 3.01$ and $\beta b < 1.39$.  It is 
first order above this tricritical point and given there by the 
thick solid line, which is the thermodynamic transition line found by comparing free 
energies. Starting in the ferromagnetic phase for $\beta J> 3.01$ and letting
$\beta b$ become bigger we arrive at the first solid line where also the 
paramagnetic solution starts to be stable and, hence, the coexistence region 
$II$ begins. This line is given by (\ref{cont}). At the thick full
line, this paramagnetic solution becomes the global
minimum of the free energy and at the second thin solid line given by
\begin{equation}
m(coth(\beta J)-m)=\frac{1}{\beta J}
\end{equation}
the ferromagnetic solution stops existing.
It is interesting to remark that inside the $m=0$ 
phase, for any $\beta J$ and $\beta b \geq 2.89$ only 10\% or less of the spins
remain in states $\pm$1  Furthermore, the phase diagram in the region of
negative $b$ is rather trivial since  negative $b$ tend to suppress all
the zero states in the system.

\begin{figure}[h]
\resizebox{0.95\columnwidth}{!}{
 \includegraphics*[angle=270, scale=0.4]{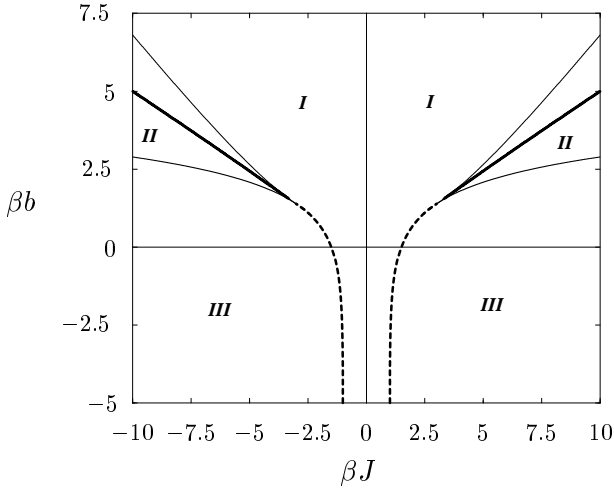}}
\caption{Phase diagram of the synchronous $Q=3$ Ising
  ferromagnet. The lines are as in fig.~1}
\label{Q=3_par_pd}       
\end{figure}

For synchronous updating the equations (\ref{parval1})-(\ref{parval2}) are  
invariant under a change of sign of $J$, such that the corresponding 
$\beta J-\beta b$ phase diagram will be symmetric with respect to the axis 
$J=0$. Furthermore, as shown before, the sequential and synchronous Q-Ising 
models have exactly the same stationary states for any $J>0$. Therefore, the 
phase diagram  is straightforwardly given in 
figure~\ref{Q=3_par_pd}. 
However, some caution is required here. From a study of the dynamics, similar 
to the one presented for the
BEG-model in the next section, one becomes aware of the difference between 
the regions $J>0$ and $J<0$. For positive $J$, all stationary solutions are 
fixed-points, while for negative $J$ the $m=0$ solution (stable in $I$ and $II$)
is of the same nature as for  $J>0$. The $m\neq 0$ solution (stable in $II$ 
and $III$) is a two-cycle, and the system jumps from $m$ to $-m$
with the activity $a$ constant. 

\section{BEG ferromagnet: statics for sequential and synchronous updating}
\label{sec4}

We recall the (pseudo)-Hamiltonian for sequential and synchronous updating
(\ref{BEGhseq}) respectively (\ref{BEGhsyn}) and choose ferromagnetic couplings 
\begin{equation}
J_{ij}=\frac{J}{N}, \quad K_{ij}=\frac{K}{N} \, .
\end{equation}
The order parameters describing the properties of the system are defined as in
(\ref{maqising}).

For sequential updating a standard calculation leads to the free energy
\begin{eqnarray}
&&{\beta}f_{S}=\underset{m,a}{\mbox{extr}}\left\{
                     \frac{1}{2}\beta Jm^2+\frac{1}{2}\beta
                       Ka^2  \right. \nonumber \\
        &&             + \left.
                     \ln{\left[\rule{0cm}{.45cm} 2\exp{(\beta Ka)}
                      \cosh{(\beta J m)}+1\right]} \right\}
\end{eqnarray}
and the following fixed-point equation must be satisfied
\begin{eqnarray} 
\label{seqm}
&& m=\frac{2\sinh{(\beta Jm)}}{\exp{(-\beta Ka)}+2\cosh{(\beta Jm)}}\\
\label{seqq}
&& a=\frac{2\cosh{(\beta Jm)}}{\exp{(-\beta Ka)}+2\cosh{(\beta Jm)}} \, .
\end{eqnarray}

For synchronous updating the free energy becomes
\begin{eqnarray}
&&{\beta}f_{P}=
   \underset{m_{\sigma},m_{\tau},a_{\sigma},a_{\tau}}{\mbox{extr}}
         \left\{\beta J m_{\sigma} m_{\tau}
                     +  \beta K a_{\sigma} a_{\tau} \right. \nonumber \\
&& \left. + \ln  \left[\left(2\exp{(\beta K a_{\sigma})}
                      \cosh{(\beta  J m_{\sigma})}+1 
                       \right) \right. \right. \nonumber \\
&& \quad\times  \left.\left.\left(2\exp{(\beta K a_{\tau})}
                      \cosh{(\beta  J m_{\tau})}+1 \right) \right] \right\}
\end{eqnarray}
with  $m_{\sigma},m_{\tau},a_{\sigma},a_{\tau} $ satisfying the saddle-point 
equations
\begin{eqnarray}
&&m_{\sigma} = F_{BEG}(m_{\tau},a_{\tau})\ , 
     \quad  m_{\tau} = F_{BEG}(m_{\sigma},a_{\sigma})\\
&&a_{\sigma} = G_{BEG}(m_{\tau},a_{\tau})\ , 
     \quad  a_{\tau} = G_{BEG}(m_{\sigma},a_{\sigma}) 
\end{eqnarray}
and the functions $F_{BEG}, G_{BEG}$ given by
\begin{eqnarray}
&& F_{BEG}(x,y)=\frac{2\sinh{(\beta Jx)}}{\exp{(-\beta Ky)}+
  2\cosh{(\beta Jx)}}  \label{funcFG1}  \\
&& G_{BEG}(x,y)=F_{BEG}(x)\coth{(\beta Jx)} \, .
\label{funcFG2} 
\end{eqnarray}

It is clear that the results for sequential updating can then be written as
\begin{equation}
\label{seqval}
m= F_B(m,a) \ , \quad  a= G_B(m,a)
\end{equation}
and the results for synchronous updating satisfy
\begin{eqnarray}
&&m_{\sigma}= F_B(F_B(m_{\sigma},a_{\sigma}),G_B(m_{\sigma},a_{\sigma}))\\ 
&&a_{\sigma}= G_B(F_B(m_{\sigma},a_{\sigma}),G_B(m_{\sigma},a_{\sigma}))
\label{parval}
\end{eqnarray}
where, for convenience,  we have simplified the subscript.

\begin{figure}[h]
\resizebox{0.95\columnwidth}{!}{
 \includegraphics*[angle=270, scale=0.6]{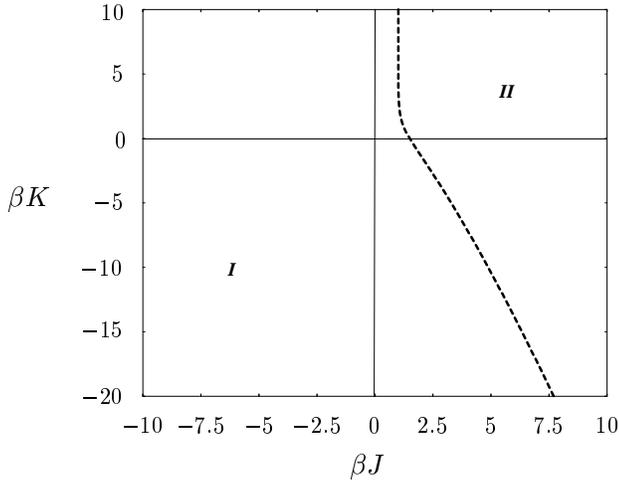}}
\caption{Phase diagram for the sequential BEG ferromagnet. The lines are as in
fig.~1}
\label{pdseq}       
\end{figure}

These relations form again the basis for studying the differences and
similarities  between sequential and synchronous updating in the BEG model. 
Let us first look at the phase diagram for sequential updating shown in
figure~\ref{pdseq}. 
For $J<0$  the only stable solution is given by $m=0$ and 
the single value $a=G_{B}(0,a)={2}{[2+\exp{(-\beta Ka)}]^{-1}}$.
For $J>0$ we see from (\ref{seqq}) that, when $m\neq 0$ and $K\neq 0$, 
$a=m\coth{(\beta Jm)}$.

The transition between the paramagnetic and ferromagnetic
phase is given by the dashed line
\begin{equation}
\label{cont2}
a=\frac{1}{\beta J},\qquad \beta K=-\beta J\ln{(2(\beta J-1))}
\end{equation}
in analogy with the transition line (\ref{cont}) for the $Q=3$ Ising 
ferromagnet.
It is  second order for all coupling parameters, in contrast with the one 
for the $Q=3$ Ising model. We remark that inside the paramagnetic
phase, only 10\% of the spins (or less) remain in the states $\pm 1$ below 
$\beta K\simeq -28.904$.

\begin{figure}[h]
\resizebox{0.95\columnwidth}{!}{
 \includegraphics*[angle=270, scale=0.65]{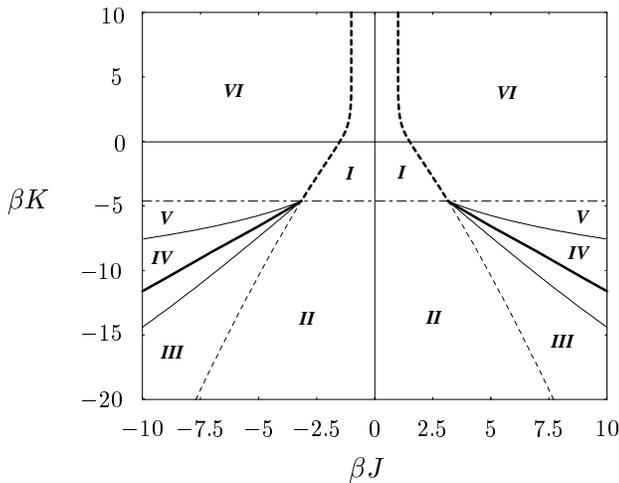}}
\caption{Phase diagram for the synchronous BEG ferromagnet. The dashed-dotted
line indicates the bifurcation of solutions in $a$. The rest of
the lines is as in fig.~1}
\label{parpd}       
\end{figure}

For synchronous updating the phase diagram is more involved as can be seen in
figure~\ref{parpd}. First we note that the set of equations (\ref{parval}) is 
invariant under the change of sign of $J$, such that the phase diagram is
symmetric with respect to the $J=0$ axis. 

\begin{figure}[h]
\resizebox{0.8\columnwidth}{!}{
 \includegraphics*[angle=270,scale=0.65]{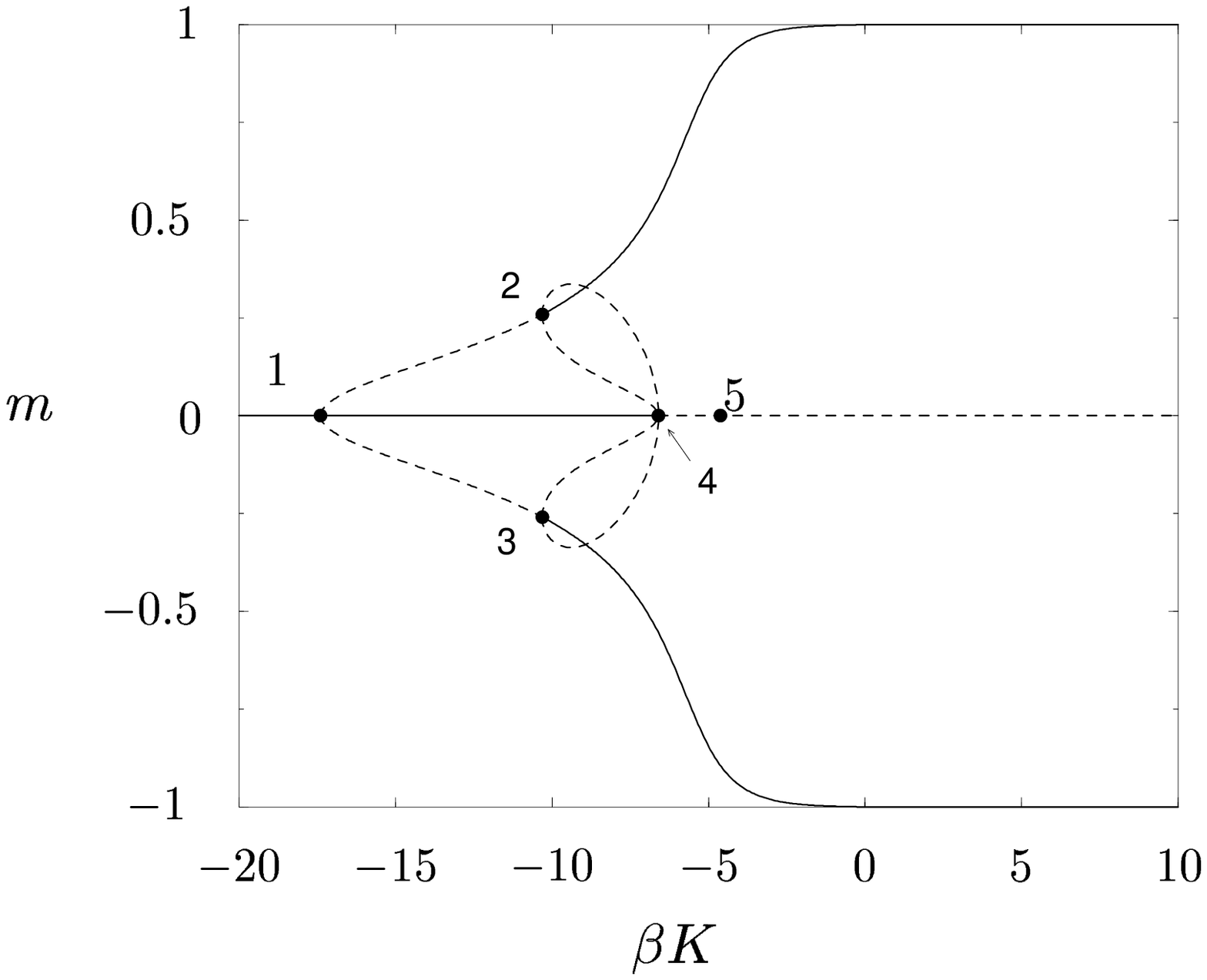}}
 \\
\resizebox{0.8\columnwidth}{!}{ 
 \includegraphics*[angle=270,scale=0.65]{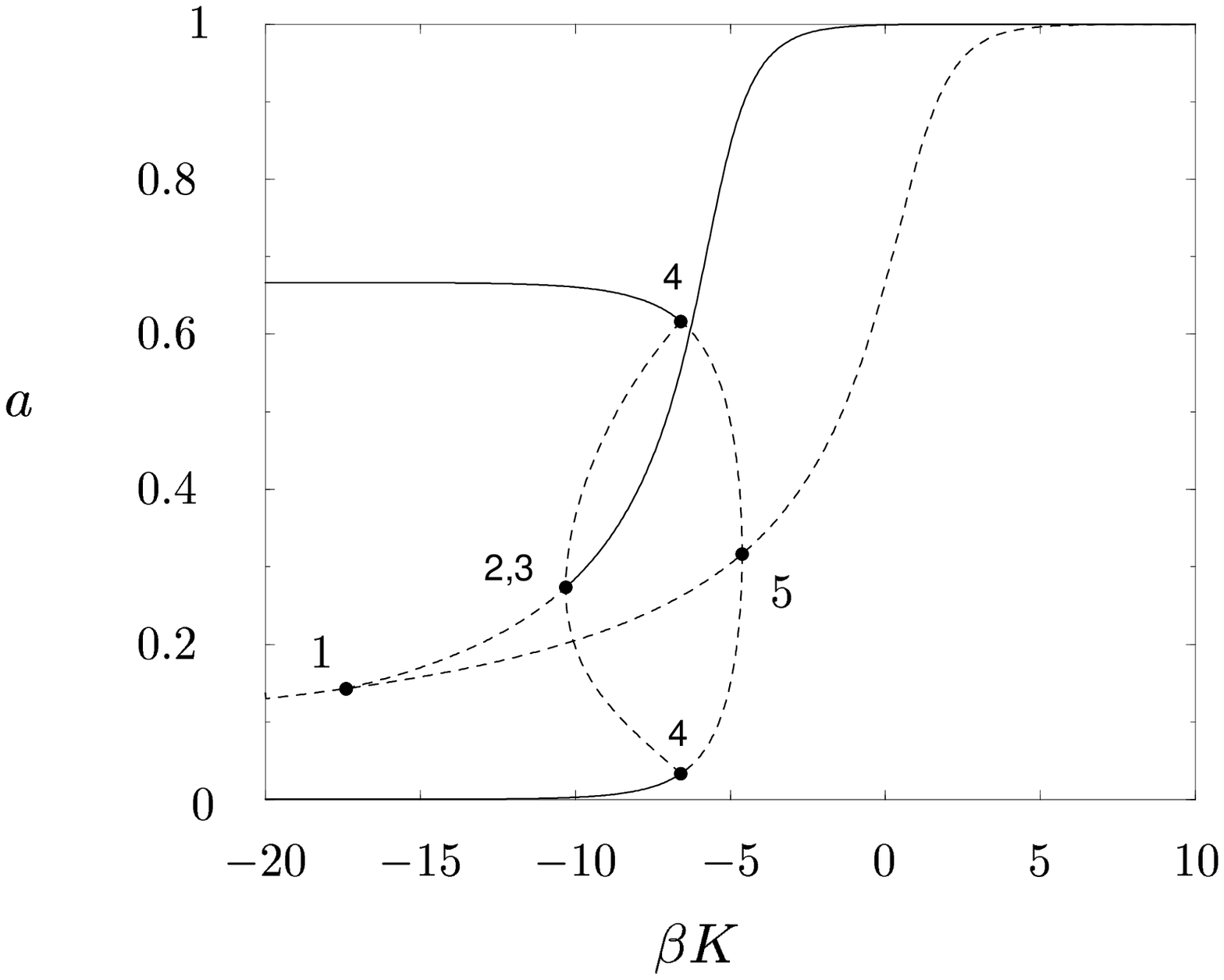}}
\caption{Stationary limit of $m$ and $a$ as a function of
  $\beta K$ given $\beta J=7$. The same labels correspond to the 
  same points. Solid lines denote stable solutions, while dashed lines 
  indicate unstable ones. }
\label{J7}       
\end{figure}

In the case $m=0$ one finds from eq. (\ref{parval}) 
that the equation for $a$  is given by 
$a=G_{B}(G_{B}(0,a))$. 
For certain values of $\beta K$, this equation has three solutions bifurcating 
from the sequential one $a=G_{BEG}(0,a)$ at the following point 
\begin{equation}
\label{need}
\beta K=\frac{-1}{a(1-a)},
    \qquad 1=(1-a)\ln{\left(\frac{2(1-a)}{a}\right)}
\end{equation}
giving the result $a^* \simeq 0.316$ and $\beta K^*\simeq -4.623$. 
This bifurcation line is indicated in fig.~\ref{parpd} as the dashed-dotted
line. It separates the regions \emph{I}-\emph{II} and \emph{V}-\emph{VI} in
the phase diagram. For $\beta K< \beta K^*$, the two new solutions appearing
at that point become automatically the stable ones in the phases
where $m=0$ is stable ($II$, $III$, and $IV$), while the sequential
solution becomes unstable. Figure~\ref{J7} illustrates this behaviour.
In region $I$, where only the solution $m=0$ is stable, the unique and stable
solution for $a$ is $a=G_B(0,a)$. In addition, for $\beta K<
\beta K^*$ the transition line  (\ref{cont2}) becomes simply
the border where the ferromagnetic solution starts to exist, but is
not yet stable, since in region $III$ the ferromagnetic solution is only a
minimum of the free energy in the $m$ direction. At this point we remark that in
figure~\ref{J7} still other spurious unstable solutions can be seen, i.e., the
loops in the top figure.

Simulations and the dynamics of the BEG-model discussed in the next section 
show that in the paramagnetic phase $II$, for $\beta K< \beta K^*$, the system
oscillates between the two stable solutions for $a$ and, therefore, the 
equilibrium
configuration is a cycle of period 2 where both configurations involved have 
different $a$ values
which become equal to $0$ and $2/3$ when $\beta K\rightarrow -\infty$.

We find that below $\beta K^*$ the
transition from the paramagnet to the ferromagnet phase becomes first order.
Comparing numerically the free energies we find the first order 
thermodynamic transition indicated by the thick solid lines in 
fig.~\ref{parpd}. The tricritical point is given by 
$\beta J \simeq\pm 3.160$, $\beta K\simeq-4.623$. As a consequence, as in the $Q=3$
Ising model and in contrast with the sequential BEG model, there is a 
coexistence
region bordered by the thin solid lines. Again, these lines can be obtained
analytically. Starting in the stable paramagnetic phase $m=0$ for, to fix 
ideas, $\beta J= 7$ and increasing the value of $\beta K$ we find the following.
We first meet the thin dashed line where the ferromagnetic solution $m>0$ 
appears but is still unstable. This line coincides with the transition line in 
the sequential BEG model (recall fig.~\ref{pdseq}) because the sequential 
solution is also a solution here. In region $III$ this ferromagnetic solution 
stays unstable until we meet the  solid line between regions $III$ and $IV$,
being the lower border of the coexistence region. This line is given by the 
point where the
ferromagnetic free energy becomes a (local) minimum in the $a$ direction, i.e.,
\begin{align}
&(2\beta K-a)\frac{\partial G_B(m,a)}{\partial a}-(\beta J)^2(a^2-m^2)
      \left(\frac{\partial F_B(m,a)}{\partial a}\right)^2 \nonumber\\
& -(1-a)\left[(\beta K)^2 a(1+\left(\frac{\partial G_B(m,a)}{\partial
            a}\right)^2  \right. \nonumber \\
        & \left.    +
      \beta K a G_B(m,a)\frac{\partial G_B(m,a)}{\partial
            a}
 -\beta J m G_B(m,a)\frac{\partial F_B(m,a)}{\partial a} \right.
                     \nonumber\\
&\left. +2\beta J \beta K
 m \frac{\partial F_B(m,a)}{\partial a}\frac{\partial G_B(m,a)}{\partial
            a}\right]=0
\end{align}
where $F_B(m,a)$ and $G_B(m,a)$ are defined in (\ref{funcFG1}) and
(\ref{funcFG2}). 

Next, we meet the thermodynamic line discussed before where the ferromagnetic
solution becomes a global minimum of the free energy. Increasing $\beta K$
further we arrive at the border of region $IV$ and $V$ where the paramagnetic
solution becomes unstable. It is given by 
\begin{equation}
\frac{2 (\beta J)^2 a}{\exp({-\beta Ka})+2}=1 \, .
\end{equation}
This equation has two solutions. The first solution $a=G_B(0,a)$ gives the
separation line between regions $I$ and $VI$, which corresponds to the
transition line in sequential updating (fig.~\ref{pdseq}). The second 
solution $a=G_B(G_B(0,a))  \neq  G_B(0,a)$ gives the upper border
of the coexistence region
$IV$. The last line we meet is the separation between regions $V$ and
$VI$, as found in eq. (\ref{need}). In regions $V$ and $VI$, only the
ferromagnetic solution is stable.

These results allow us to say that sequential and synchronous updating lead to
completely the same physics ($f_P=2f_S$) in the region 
$\beta J>0$, $\beta K>-4.623$.
Hence, as we will further explain in the next section on dynamics there are no
cycles for positive couplings, but we do find them for positive $\beta J$
and negative $\beta K$. They turn out to be stable in the regions 
$II$, $III$ and $IV$. 

\section{BEG ferromagnet: dynamics for sequential and synchronous updating}
\label{sec5}

The aim of this section is to study the dynamics of the BEG model in order to
further examine the difference between sequential and synchronous updating and
to further understand the appearance and behaviour of two-cycles. In order to do
so we use the generating function (path integral) technique introduced in 
\cite{MSR73} to the field of statistical mechanics and, by now, part of many 
textbooks. In particular, we follow \cite{C01d}. Since we  have no disorder 
in our problem, the method can be used in its simplest form. 

The probability of a certain microscopic path of spin configurations from time
$0$ up to time $t$ is denoted by  $\mbox{P} [\bsigma(0),...,\bsigma(t)]$. 
For the BEG model defined in section~\ref{sec2} it is given by 
\begin{equation}
\mbox{P}[\bsigma(0),...,\bsigma(t)]\equiv
\mbox{P}_0(\bsigma(0)) \prod_{s=0}^{t-1}
W[\bsigma(s+1);\bsigma(s)]
\end{equation}
with $W[\bsigma;\bsigma']$ the transition matrix of the Markovian process
defined by the spin-flip dynamics given by eqs.~(\ref{eq:trans}) and
(\ref{potbeg}). It depends on the specific way of updating the spins
(sequential or synchronous) and will be specified later. 
We introduce a generating function for the BEG model as a function of the 
field $\bf \Phi$ 
\begin{equation}
Z[{\bf \Phi}]=\left\langle \exp\left[-i\sum_{s=0}^t\sum_{i=1}^N
\sum_{k=1}^2\phi_{k,i}(s)\sigma_i^k(s)\right]\right\rangle_{path}
\label{genfun}
\end{equation}
where the average $\langle\cdot\rangle_{path}$ is an average over 
$\mbox{P} [\bsigma(0),...,\bsigma(t)]$. The order parameters of the system
are generated by this function $Z[{\bf \Phi}]$ through
\begin{equation}
\langle\sigma_i^k(s)\rangle_{path}=i\lim_{{\bf \Phi}\rightarrow
  0}\left(\frac{\partial Z[{\bf \Phi}]}{\partial \phi_{k,i}(s)}\right) \, .
\end{equation}
At this point we remark that, for our purposes, we only look at these one-time
quantities. Again, to unify notation we use the ``magnetizations''
$m_k (s)$, $k=1,2$ to denote the magnetization $m(\bsigma(s))$, respectively the
spin activity $a(\bsigma(s))$.

Introducing these magnetizations into the generating function (\ref{genfun}) by
using appropriate $\delta$ functions and grouping the terms in those depending
on the site index and those which do not, we obtain
\begin{equation}
\label{start1}
Z[{\bf \Phi}]\propto\int\prod_k\left[d{\bf m}_kd\hat{{\bf m}}_k\right]\exp{N\Psi}
\end{equation}
with $d{\bf m}_k=dm_k(0)...dm_k(t)$ and similarly for $d\hat{{\bf m}}_k$. The
quantity $\Psi$ reads
\begin{eqnarray}
\label{start2}
&&\Psi=i\sum_{k,s}\hat{m}_k(s)m_k(s)  \nonumber \\
 && + \frac{1}{N}\sum_i\ln{\left\langle \exp \left[-i\sum_{k,s}
   \sigma_i^k(s)(\hat{m}_k(s)+\phi_{k,i}(s))\right]\right\rangle_{path}}
\end{eqnarray}
This generating function (\ref{start1}) allows for the application of the 
saddle-point method. In order to continue we need to specify the type of 
updating.

\subsection{Synchronous updating}

In this case all spins are updated at the same time such that the transition
probabilities $W_p[\bsigma(s+1);\bsigma(s)]$ are just the product of the 
transition probabilities of the single spin (recall eqs.~(\ref{eq:trans})
and (\ref{potbeg})). Noting that the local fields are equal to 
$h_k(s)\equiv J_k m_k(s)$ (with obvious definitions for $J_k$) we obtain
\begin{eqnarray}
&&\Psi=i\sum_{k,s}\hat{m}_k(s)m_k(s)
   +\frac{1}{N}\sum_i\ln\left[\sum_{\bsigma(0)}...\sum_{\bsigma(t)}
              \mbox{P}_0(\bsigma(0)) \right. \nonumber\\
&& \left. \exp{[-i\sum_{k,s} \sigma_i^k(s)(\hat{m}_k(s)+\phi_{k,i}(s)+i\beta
        J_km_k(s-1))]} \right.  \nonumber\\
&&  \left. \exp{\bigg[-\sum_s\ln{\left(
     1+2e^{\beta J_2m_2(s-1)}\cosh{(\beta J_1m_1(s-1))}\right)}\bigg]}\right]
     \nonumber \\
\end{eqnarray}
Choosing the initial conditions $\mbox{P}_0(\bsigma(0))$ to be iidrv with 
respect
to $i$ and letting $\phi_{k,i} \rightarrow \phi_k$, the single-site nature of the
last expression becomes apparent. Defining an effective (i.e., single-site) path
average denoted
by $\langle\cdot\rangle_{*}$  the saddle-point equations then become
\begin{equation}
m_k(s)=
    \langle\sigma^k(s)\rangle_* \, ,\quad
\hat{m}_k(s)=0\,.
\label{saddlemmhat}
\end{equation}
Working out further details and summing over the spins we easily obtain
\begin{align}
 m_1(s)&=
  \frac{2e^{\beta J_2m_2(s-1)}\sinh{(\beta J_1m_1(s-1))}}
    {1+e^{\beta J_2m_2(s-1)}\cosh{(\beta J_1m_1(s-1))}}\\
m_2(s)&=
m_1(s)\coth(\beta J_1m_1(s-1)) \, .
 \end{align}
These saddle-point equations allow for two-cycles when 
$m_k(s)=m_k(s+2)$, $k=1,2$ and  fixed points with $m_k(s)=m_k(s+1)$. The
stationary limit is  obtained when we drop the time dependence,
writing $m_k(s)$ as $m_{\sigma}$ and $m_k(s-1)$ as $m_{\tau}$ or
the other way around. Some further discussion and numerical results will be 
presented after studying sequential updating.

\subsection{Sequential updating}

We start from the stochastic process
\begin{equation}
\label{put}
p_{s+1}(\bsigma)=\sum_{\bsigma'}W_{s}[\bsigma;\bsigma']\,p_s(\bsigma')
\end{equation}
with $p_{s+1}(\bsigma)$ the probability to be in a state $\bsigma$ at
time $s+1$. For the BEG
model 
\begin{eqnarray}
\label{put2}
W_{s}[\bsigma;\bsigma']&=&
  \frac{1}{N}\sum_i\left\{
  w_i(\bsigma)\delta_{\bsigma,\bsigma'} \right. \nonumber\\
  &+& \left. w_i(F_i\bsigma)\delta_{\bsigma,G_i\bsigma'}+
  w_i(G_i\bsigma)\delta_{\bsigma,F_i\bsigma'}\right\}
\end{eqnarray}
with the shorthand 
$w_i(\bsigma)\equiv\mbox{P}\{\sigma_i(s+1)=\sigma_i|\bsigma(s)\}$ and where
$F_i$ and $G_i$ are cyclic spin-flip operators between the spin 
states \{-1,0,+1\}  defined by
\begin{eqnarray}
&&F_i\Phi(\bsigma)
    =\Phi(\sigma_1,...,\sigma_{i-1},\frac{-3\sigma_i^2-\sigma_i+2}{2},
       \sigma_{i+1},...,\sigma_N) \nonumber \\
&& G_i\Phi(\bsigma)= F_i(F_i\Phi(\bsigma)) \, .
\end{eqnarray}
Each time step a randomly chosen spin is updated. In the thermodynamic limit the
dynamics becomes continuous because the characteristic time scale is $N^{-1}$.
The standard procedure is then to  update a random spin according to
(\ref{eq:trans}) and (\ref{potbeg}) with time intervals $\Delta$ that are 
Poisson distributed with mean $N^{-1}$ \cite{BLLS71}. 
We can then write a continuous master equation in the thermodynamic limit
\begin{eqnarray}
\frac{d}{ds}p_s(\bsigma)&\equiv& \lim_{\Delta\rightarrow
  0}\frac{p_{s+\Delta}(\bsigma)-p_s(\bsigma)}{\Delta} \nonumber \\
&=&\sum_i\left\{
(w_i(\bsigma)-1)p_s(\bsigma) \right. \nonumber \\
&+& \left.  w_i(F_i\bsigma)p_s(F_i\bsigma)+
    w_i(G_i\bsigma)p_s(G_i\bsigma)\right\}
    \label{conp}
\end{eqnarray}

Starting again from the generating function (\ref{start1})-(\ref{start2}),
the average over the paths has to be understood as an average over a 
constrained process given by eqs. (\ref{put})-(\ref{put2}) in which the 
overlaps are prescribed at all time steps. Therefore, due to the introduction
of the ${\bf m}_k(s)$ and $\hat{{\bf m}}_k(s)$, the transition
probabilities should be written as a function of these overlaps.
$w_i(\bsigma(s)) \rightarrow  w_i(m_1(s),m_2(s))$. 
The key step is to write this stochastic process as a single-site problem. 
This is possible when noting that the $p_s(\bsigma)$ can be written as
\begin{equation}
p_s(\bsigma)=\prod_{i=1}^N\left[1-\sigma_i^2+
   \frac{\sigma_i}{2}\tilde{m}_{1i}(s)
+\left(\frac{3\sigma_i^2}{2}-1\right)\tilde{m}_{2i}(s)\right]
\end{equation}
where, in order to satisfy eq.~(\ref{conp}) the 
$\tilde{m}_{k,i}(s)\equiv \langle\sigma_i^k(s)\rangle_{path}$ have to obey the
following evolution equations
\begin{eqnarray}
&&\frac{d}{dt}\tilde{m}_{1,i}(s) \,\,\,\,\, \nonumber \\
&& = \frac{2e^{\beta J_2m_2(s)}
         \sinh{(\beta J_1m_1(s))}}{1+2e^{\beta
      J_2m_2(s)}\cosh{(\beta J_1m_1(s))}}-\tilde{m}_{1,i}(s) \,\,\,\,\, \\
&&\frac{d}{dt}\tilde{m}_{2,i}(s) \nonumber \\  
&&   =\frac{2e^{\beta J_2m_2(s)}\cosh{(\beta J_1m_1(s))}}
   {1+2e^{\beta J_2m_2(s)}\cosh{(\beta J_1m_1(s))}}-\tilde{m}_{2,i}(s)\,\,
\end{eqnarray}
with the initial conditions $p_0(\bsigma)=\delta_{\bsigma,\bsigma(0)}$.
These evolution equations are clearly site independent.

The  function $\Psi$ for continuous time then reads
\begin{eqnarray}
&&\Psi=i\sum_k\int
ds\,\hat{m}_k(s)m_k(s) \nonumber \\
&& \hspace*{-0.3cm} +\frac{1}{N}\sum_i\ln{\left\langle
    \exp \left[{-i\sum_k\int ds\,\sigma_i^k(s)(\hat{m}_k(s)+\phi_{k,i}(s))}
      \right]\right\rangle_{path}} \nonumber \\
\end{eqnarray}
and by choosing the initial conditions iidrv with respect to $i$ and letting 
$\phi_{k,i} \rightarrow \phi_{k}$ the single-site nature is complete.
Defining an effective path average denoted, as before, by 
$\langle\cdot\rangle_{*}$, the saddle-point equations are formally the same as 
eqs.~(\ref{saddlemmhat}) implying that
$m_k(s)=\tilde{m}_{ki}(s),\,\forall i$. Hence, the
final evolution equations for the order parameters of the BEG model with
sequential updating are
\begin{eqnarray}
&&\frac{d}{dt}{m}_{1}(s) \,\,\,\,\, \nonumber \\
&& = \frac{2e^{\beta J_2m_2(s)}
         \sinh{(\beta J_1m_1(s))}}{1+2e^{\beta
      J_2m_2(s)}\cosh{(\beta J_1m_1(s))}}-{m}_{1}(s) \,\,\,\,\, \\
&&\frac{d}{dt}{m}_{2}(s) \nonumber \\  
&&   =\frac{2e^{\beta J_2m_2(s)}\cosh{(\beta J_1m_1(s))}}
   {1+2e^{\beta J_2m_2(s)}\cosh{(\beta J_1m_1(s))}}-{m}_{2}(s)\,\,
\end{eqnarray}
Clearly, the stationary solutions obtained by forgetting about the time
dependence do not allow two-cycles.

\subsection{Simulations and numerical results}
We illustrate our findings by showing some numerical results and comparing
these with simulations for up to $N=500000$ spins. 
Recall figs.~\ref{pdseq} and \ref{parpd}. For some typical values of the 
couplings in the ferromagnetic phase, e.g., $\beta J_1=\beta J=3$, 
$\beta J_2=\beta K=1$ in region $VI$, both types 
of updating lead to the same stationary state, with $m_1\simeq
m_2\simeq 1$, the only difference being the
speed with which this happens: sequential updating seems to be a bit slower.  
For $\beta J=-3$ and $\beta K=1$ sequential updating leads to
$m_1=m=0$, $m_2=a\simeq 0.8$
while synchronous updating  gives a cycle in $m$ with $|m|\simeq a \simeq 1$.
Simulations for these cases are in excellent agreement with these results.
\begin{figure}[h]
\resizebox{0.95\columnwidth}{!}{
 \includegraphics*[angle=270,scale=0.4]{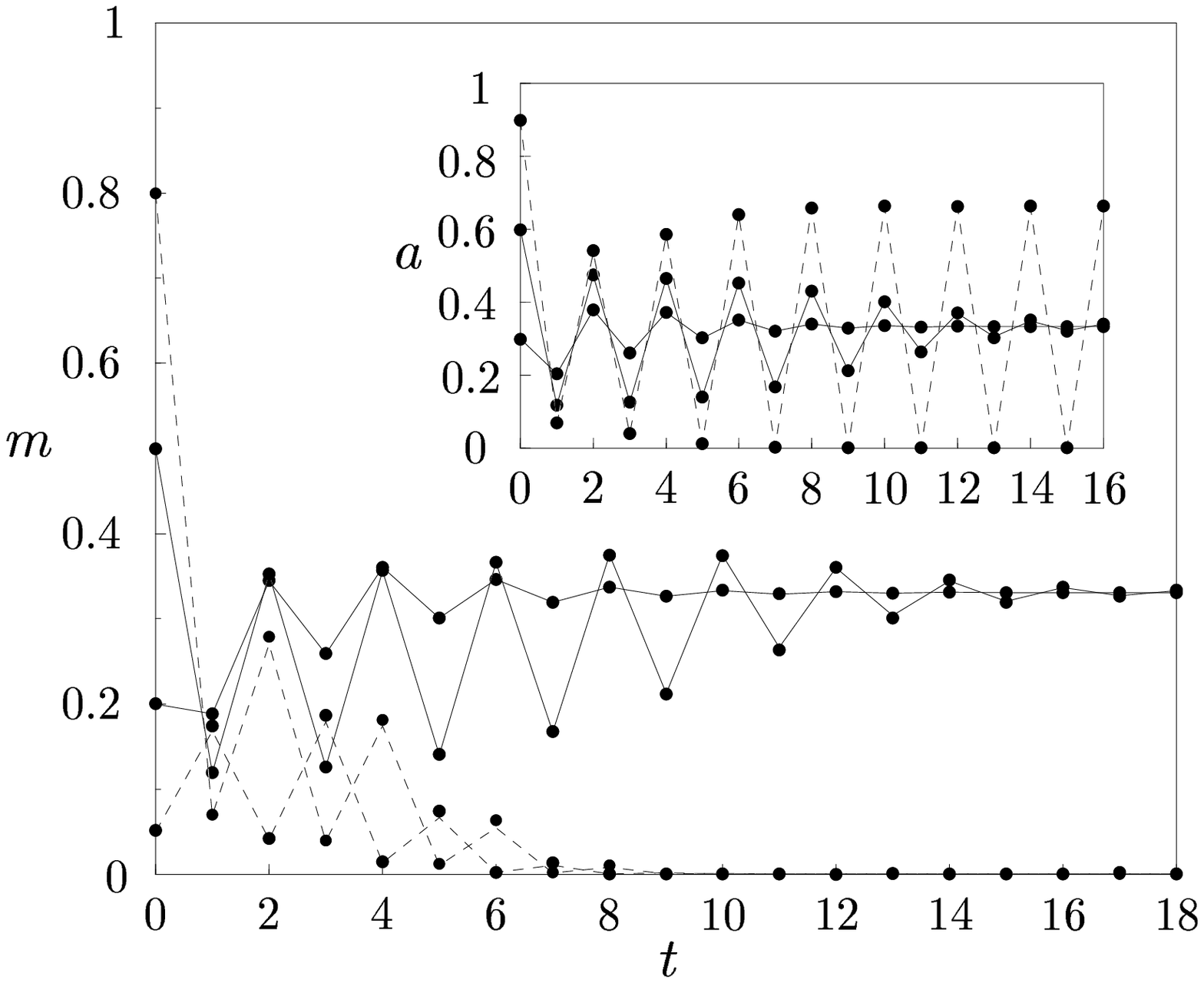}}
 \\
\resizebox{0.95\columnwidth}{!}{ 
 \includegraphics*[angle=270,scale=0.4]{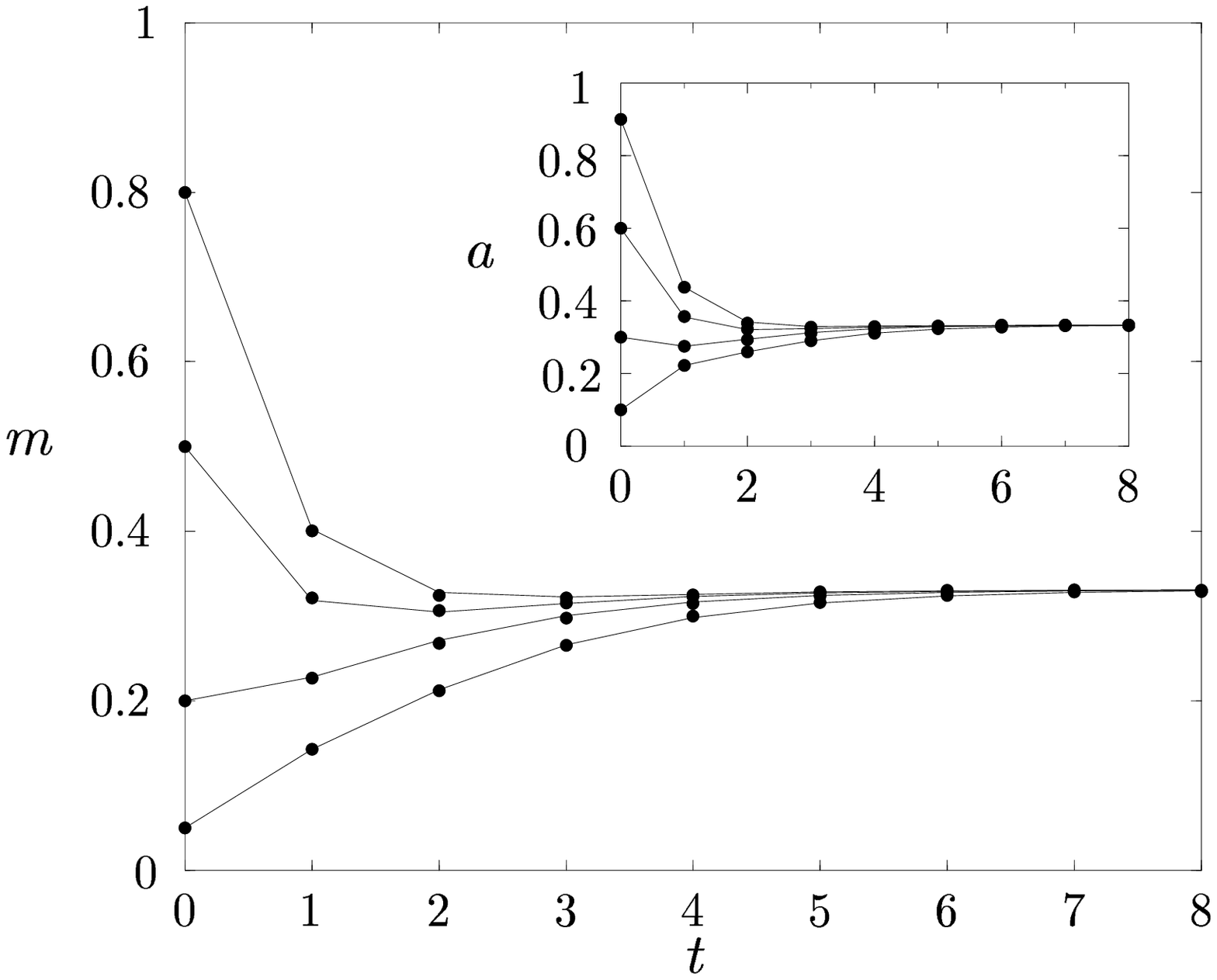}}
\caption{Evolution of $m$ and $a$ as a function of time for sequential
(bottom) and synchronous (top) updating for $\beta J=8$, $\beta K=-10$. The dots
correspond to simulation points. For synchronous dynamics, the paths
leading to a cycle in $a$ have been plotted with dashed lines.}
\label{two}       
\end{figure}
\begin{figure}[h]
\resizebox{0.95\columnwidth}{!}{
 \includegraphics*[angle=270,scale=0.4]{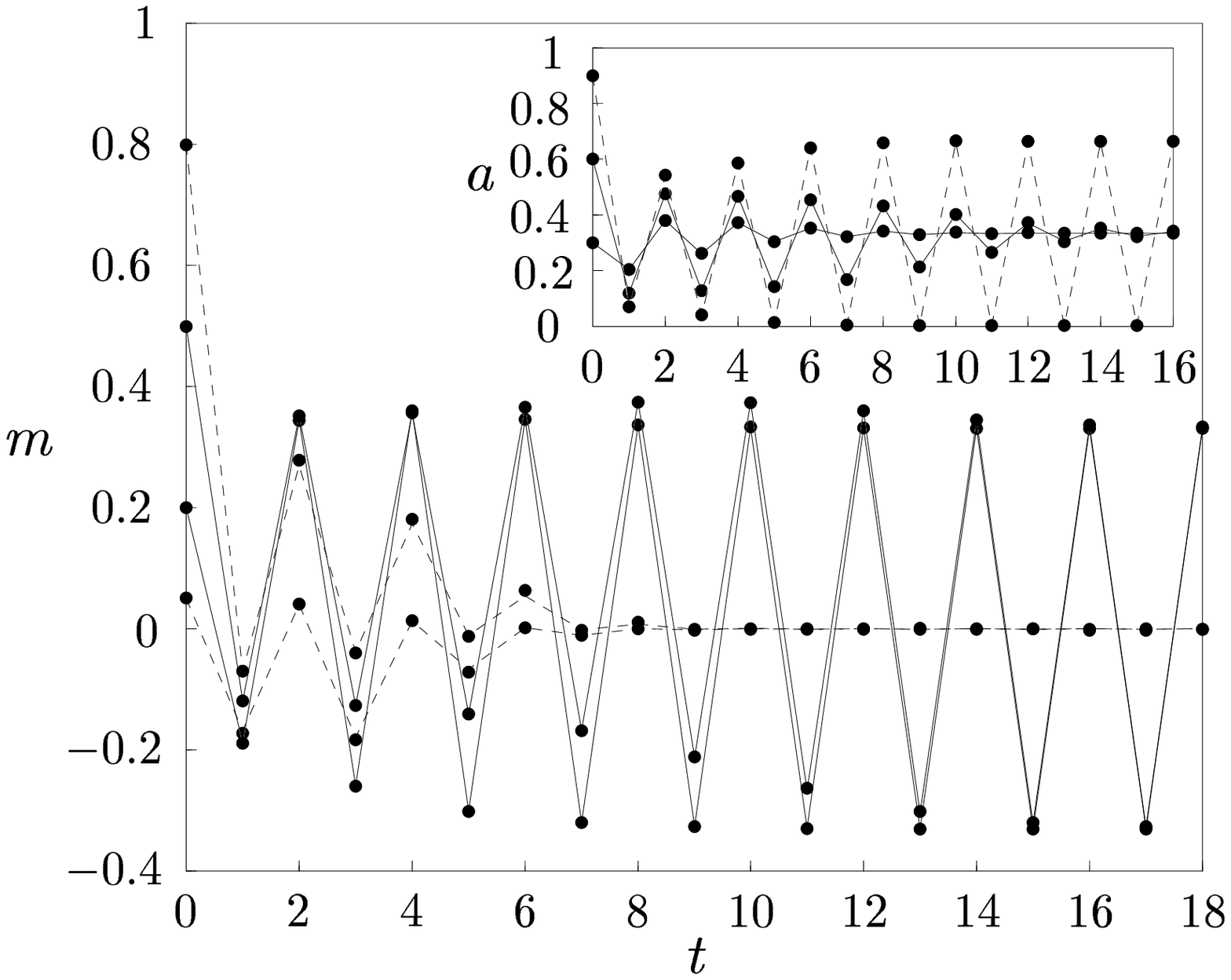}}
 \\
\resizebox{0.95\columnwidth}{!}{ 
 \includegraphics*[angle=270,scale=0.4]{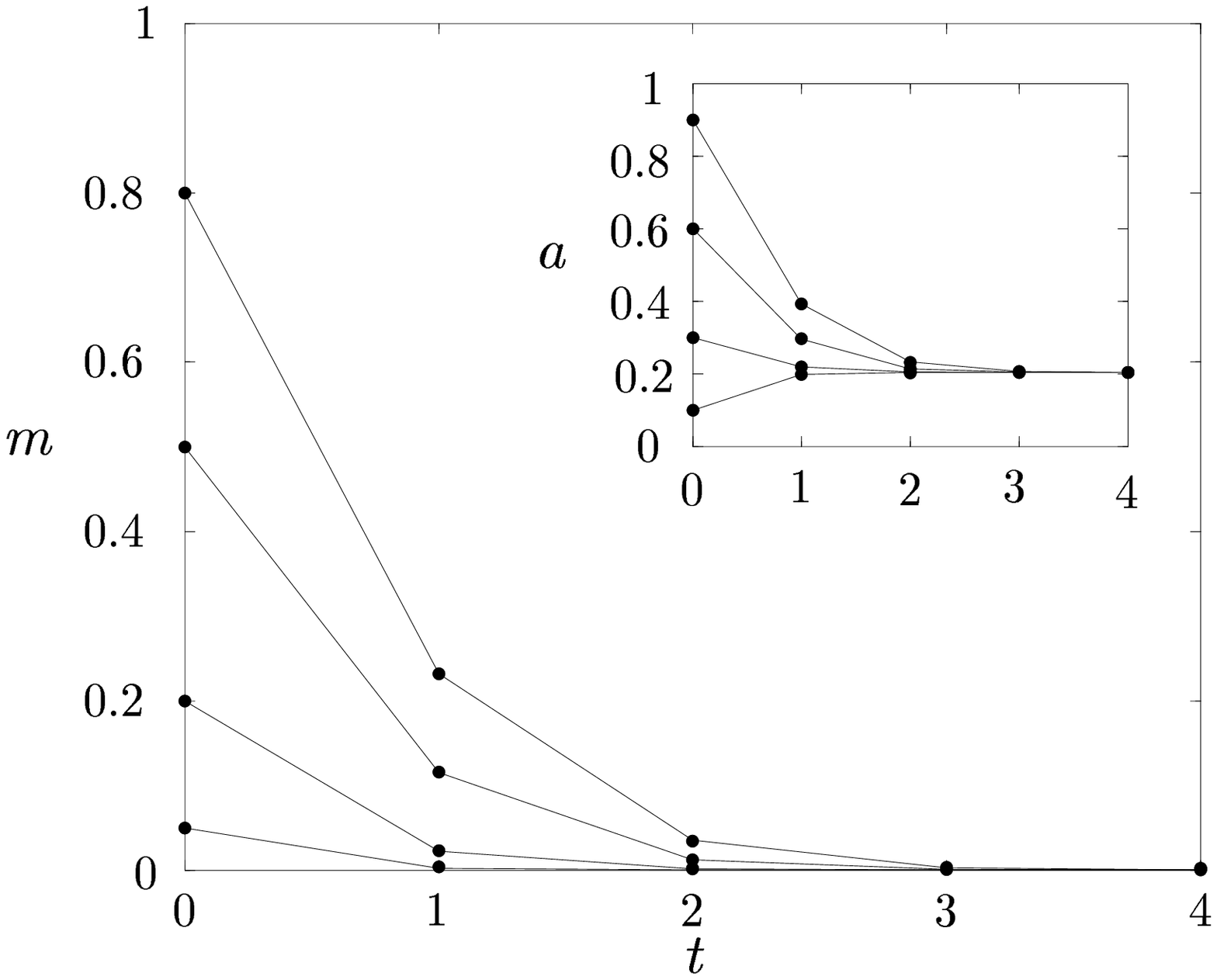}}
\caption{Evolution of $m$ and $a$ as a function of time for sequential
(bottom) and synchronous (top) updating for $\beta J=-8$, $\beta K=-10$. The dots
correspond to simulation points. For synchronous dynamics, the paths
leading to a cycle in $a$ have been plotted with dashed lines.}
\label{three}       
\end{figure}

The second set of points lie in region IV of the phase diagram fig.~4, i.e., 
$\beta J=\pm 8$ and $\beta K=-10$ and the results
of the dynamics are shown in figs.~\ref{two} and \ref{three}. The dots
correspond to simulations points. When $\beta J >0$ (fig.~\ref{two}) we see 
that the sequential system (bottom) always goes to the ferromagnetic solution
for any initial condition. We note that for sequential dynamics, $t=1$
corresponds to 1 update per spin in average. 
The synchronous system (top), however,  has two minima in the free energy, and 
depending on the initial condition it evolves to the $m=0$ solution or to 
the $m>0$ ferromagnetic one.
In addition, the basin of attraction is somewhat
involved in the sense that  the initial conditions $m(0)=0.8$ ($a=0.9$) and
$m(0)=0.05$ ($a=0.1$) lead to the $m=0$ solution, while the
initial conditions $m(0)=0.5$ ($a=0.6$) and
$m(0)=0.2$ ($a=0.3$) lead to the ferromagnetic one. The behaviour
in $a$ is as expected: when $m$ reaches the ferromagnetic
solution, $a$ tends to a single finite value, while when $m=0$,
$a$ enters a two-cycle. Indeed we are in the region of the phase diagram 
where three solutions for $a$ are allowed (recall fig.~\ref{parpd}). 
For the sake of clarity, we have only included one of the cycles in $a$ 
for the synchronous updating figures (the one for $m(0)=0.8$, $a(0)=0.9$)

When $\beta J<0$ (fig.~\ref{three}) the sequential system
always evolves to the $m=0$ solution, while the synchronous one shows a 
similar behaviour as in fig.~\ref{two}, the only difference being that now the
ferromagnetic solution is a two-cycle in $m$.

\section{Concluding remarks}
\label{sec6}
In this paper we have studied some of the physical consequences of the way 
spins are updated, sequentially respectively synchronously, in classical 
multi-state Ising-type spin systems.
First, we have derived the general form of the (pseudo-) Hamiltonian for 
$Q$-Ising and Blume-Emery- Griffiths (BEG) spin-glasses with synchronous 
updating on the basis of detailed balance.

Next, in order to study the precise differences in the stationary behaviour 
we have chosen to simplify these models to $Q$-Ising and BEG (anti-)
ferromagnets, on the one hand because these are exactly solvable both through
a free-energy analysis and a functional integration approach and on the other
hand because we did not find these results in the literature.

In the case of the $Q=3$ Ising model, no surprising behaviour has been found in
the sense that the phase diagram for synchronous updating is symmetric with 
respect to the zero-coupling axis $J=0$, and that the same stationary
solutions appear as for sequential updating except for negative couplings where
cycles of period two in $m$ occur in the ferromagnetic phase.

The differences in the behaviour of the BEG (anti)-ferromagnet are partly
unexpected. Whereas the phase diagram for sequential updating is even simpler
than the corresponding one for the $Q=3$ Ising model, a much richer phase 
diagram appears in the case of synchronous updating. 
Symmetry with respect to the axis $J=0$ still persists in this case, but 
the presence of a second relevant order parameter allows for
much richer behaviour. The region of negative $K$ coupling is characterized by
a more complicated free energy landscape. For instance, when $\beta K$
is sufficiently negative three  paramagnetic solutions exist with different
values for the spin activity, two-cycles in $a$ appear in different regions of
the parameter space and a coexistence region 
of the ferromagnetic and paramagnetic solutions is found for certain values of
the coupling parameters. 
When looking at the Hamiltonian of the BEG system, one 
expects the most interesting behaviour in the region $J>0$,
$K<0$ (also $J<0$ for synchronous dynamics), since both terms in the
Hamiltonian favour different states for the spins. Moreover,
the fact that for synchronous dynamics one has to work with two types of 
spins makes the picture still more involved. 

These findings suggest that also in more complicated disordered spin systems
like the BEG spin-glass or neural network the differences between 
sequential and synchronous updating might be much richer and more interesting 
than one expects.

\section*{Acknowledgments}
The authors thank Nikos Skantzos for informative and critical
discussions. This work has been supported in part by the Fund of Scientific
Research, Flanders-Belgium.


\end{document}